\documentclass[12pt]{iopart}

\usepackage{amsgen}
\usepackage{amssymb}
\usepackage{setstack}
\usepackage{bbold}
\usepackage{tikz}
\usetikzlibrary{decorations.pathmorphing,calc}
\usetikzlibrary{decorations.pathmorphing}
\tikzset{font={\fontsize{22pt}{12}\selectfont}}
\usepackage{multirow}
\usepackage{subcaption}
\usepackage{graphicx}
\usepackage[breaklinks=true,colorlinks=true,linkcolor=blue,urlcolor=blue,citecolor=blue]{hyperref}

\def\be{\begin{equation}}
\def\ee{\end{equation}}
\def\bea{\begin{eqnarray}}
\def\eea{\end{eqnarray}}

\begin{document}

\title{Exact Initial Data for Black Hole Universes with a Cosmological Constant}

\author{Jessie Durk$^1$ and Timothy Clifton$^2$}
\address{School of Physics and Astronomy, Queen Mary University of London, UK}
\ead{$^1$j.durk@qmul.ac.uk, $^2$t.clifton@qmul.ac.uk}

%\pacs{98.80.Jk, 04.20.Jb}

\begin{abstract}
We construct exact initial data for closed cosmological models filled with regularly arranged black holes in the presence of $\Lambda$. The intrinsic geometry of the 3-dimensional space described by this data is a sum of simple closed-form expressions, while the extrinsic curvature is just proportional to $\Lambda$. We determine the mass of each of the black holes in this space by performing a limiting procedure around the location of each of the black holes, and then compare the result to an appropriate slice through the Schwarzschild-de Sitter spacetime. The consequences of the inhomogeneity of this model for the large-scale expansion of space are then found by comparing the lengths of curves in the cosmological region to similar curves in a suitably chosen Friedmann-Lemaitre-Robertson-Walker (FLRW) solution. Finally, we locate the positions of the apparent horizons of the black holes, and determine the extremal values of their mass, for every possible regular arrangement of masses. We find that as the number of black holes is increased, the large-scale expansion of space approaches that of an FLRW model filled with dust and $\Lambda$, and that the extremal values of the black hole masses approaches that of the Schwarzschild-de Sitter solution.
\end{abstract}

\section{Introduction}

Cosmological models filled with lattices of black holes have received considerable attention over recent years, as they provide an inhomogeneous alternative to the spatially homogeneous and isotropic Friedmann-Lema\^{i}tre-Robertson-Walker (FLRW) solutions, while remaining simple enough to solve using analytic methods. They therefore present us with an opportunity to investigate some of the foundational assumptions that go into the standard concordance model of cosmology, which both neglects the effects of inhomogeneity on the large-scale expansion \cite{wiltshire}, whilst simultaneously requiring the existence of a ``dark sector'' \cite{linder}. This is of particular interest as it is theoretically possible for small-scale inhomogeneities to play a role in the large-scale dynamics of cosmological models \cite{buchert,chris}, and because inhomogeneities can have consequences for a wide array of cosmological observables \cite{kras}. It is important to ensure that we understand all of these effects, especially with the advent of precision cosmology and in anticipation of future surveys. Black hole lattices provide a valuable contribution to the literature on this subject, as they all include all relativistic effects at all orders.

The work in this paper builds on the seminal works by Lindquist and Wheeler \cite{lind} and Misner \cite{mis}, who considered time-symmetric initial data for black holes in vacuum. This work was recently extended and investigated within a cosmological context \cite{tim2}, and the evolution of this space has now been studied using numerical and symmetry-reducing methods \cite{bent}-\cite{ev5}. Perturbative and numerical analyses of cubic lattices of black holes have also been performed \cite{brune}-\cite{vij}. The present study contributes to this literature by constructing exact initial data for a universe that contains both black holes and a cosmological constant, $\Lambda$. These reduce to the models in \cite{tim2} in the limit where $\Lambda$ vanishes. These models do not require any approximations or perturbative methods, and therefore include all relativistic effects. This means that while the distribution of matter is highly idealised, the treatment of the gravitational field is exact and unambiguous.

The lattice models we consider for the majority of this paper are based on 3-spheres tessellated with regular polyhedra, with a Schwarzschild-de Sitter-like mass placed at the centre of each. They therefore represent closed cosmological models, and have a degree of homogeneity if coarse-grained over large enough scales. There are six different ways of constructing such models, with either $5$, $8$, $16$, $24$, $120$ or $600$ masses. In the absence of $\Lambda$, each of these geometries is a time-symmetric hypersurface with vanishing extrinsic curvature, corresponding to a closed FLRW universe at its maximum of expansion. With the inclusion of a non-zero $\Lambda$ we find that solutions that correspond to spaces of constant mean extrinsic curvature have the terms involving Weyl curvature and spatial curvature cancel each other out in the Hamiltonian constraint equation. In a Friedmann cosmology, this would correspond to the moment when the density parameters for dust and spatial curvature are equal in magnitude, and opposite in sign. Of course, modern observational data constrains the real universe to be very close to spatially flat. The initial data we present should therefore be considered as a mathematical exploration of exact cosmological solutions to the Einstein field equations, rather than a model to be immediately applied to the actual universe. We expect the knowledge gained from this study to be instructive for understanding inhomogeneity in more realistic models, which contain both dark matter and baryons, even though it is itself highly idealized.

Recent related works of particular significance are the numerical studies of Yoo {\it et al} \cite{yoo}, and perturbative constructions of Sanghai {\it et al} \cite{vij}. The first of these constructs a flat, infinite, cubic lattice of black holes with a non-zero cosmological constant, and numerically solves the constraint and evolution equations for the spacetime. The second study extends the post-Newtonian formalism to include a cosmological constant, and then joins together cells of perturbed Minkowski spacetime using the Israel junction conditions. Both of these studies are motivated by the fact that a positive-valued cosmological constant causes accelerated expansion of the universe at late times, and so is a prime candidate for ``dark energy''. Our work differs from these previous studies as we make no approximations when modelling the geometry of space, and because we solve the constraint equations using a fully analytic approach. The topology of our model is also different to that of Yoo {\it et al}, which allows us the benefit of studying six different tessellations (rather than the one tessellation that exists in flat space).

This paper is organised as follows. In Section  \ref{sec:solution} we present the initial data problem for a universe containing black holes and a cosmological constant. In Section \ref{sec:proper} we determine the mass of these black holes by looking for a particular constant mean curvature slice of the Schwarzschild-de Sitter spacetime. Section \ref{sec:decel} introduces the cosmological deceleration parameter, and presents results for how it is affected by back-reaction in our inhomogeneous models. In Section \ref{sec:locations} we derive a formula for locating apparent horizons within each of our lattice models, which are used in Sections \ref{sec:distances} to determine the distance between black holes. Section \ref{sec:extremal} then contains a presentation of the extremal values of $\Lambda$ in cosmologies of this type. Finally, we discuss our results in Section \ref{sec:discuss}. Throughout this paper we use geometrised units where $c = G = 1$. We use the second half of the Greek alphabet ($\mu, \nu$) as indices to denote spacetime coordinates, whilst the second half of the Latin alphabet $(i, j)$ are used as indices to denote spatial coordinates.

\section{Exact Initial Data with ${\mathbf \Lambda}$}
\label{sec:solution}

The $3 + 1$ decomposition of Einstein's equations yields the Hamiltonian and momentum constraint equations,
\begin{eqnarray}
\label{1}
     \mathcal{R} + K^2 -K_{ij}K^{ij} = 16 \pi \rho \, , \\
     D_j(K_i^{\,\,j} - \gamma_i^{\,\,j}K) = 0
     \label{2}
\end{eqnarray}
where $\mathcal{R}$ is the Ricci scalar of the hypersurface, $K_{ij}$ is the extrinsic curvature, $K$ its trace, $\rho$ is the matter density, $\gamma_{ij}$ is the intrinsic metric of the hypersurface and $D_j$ is the covariant derivative with respect to this metric. We can then write the extrinsic curvature in terms of its trace and trace-free part, $A_{ij}$, as
\begin{equation}
\label{k1}
     K_{ij} = \frac{1}{3}\gamma_{ij}K +A_{ij} \, ,
\end{equation}
and performing a conformal rescaling of the 3-metric, such that $\gamma_{ij}=\psi^4 \tilde{\gamma}_{ij}$, then allows us to write the Ricci scalar as
\begin{equation}
\label{r1}
     \mathcal{R} = \psi^{-4}\tilde{\mathcal{R}} - 8 \psi^{-5}\tilde{D}^2\psi \, ,
\end{equation}
where $\tilde{\mathcal{R}}$ is the Ricci scalar of $\tilde{\gamma}_{ij}$, and $\tilde{D}^2$ is the covariant Laplacian associated with $\tilde{\gamma}_{ij}$.

Substituting Eqs. (\ref{k1}) and (\ref{r1}) into Eqs. (\ref{1}) and (\ref{2}) then gives 
\begin{eqnarray}
\label{5}
    8\tilde{D}^2 \psi - \psi \tilde{\mathcal{R}} - \frac{2}{3}\psi^5 K^2 + \psi^{5}A_{ij}A^{ij}= -2\Lambda \psi^5,\\ 
      D_j \left( A_i^{\,\,j} - \frac{2}{3}\gamma_i^{\,\,j}K \right)= 0 \, ,
      \label{6}
\end{eqnarray}
where we have made the substitution $8 \pi \rho = \Lambda$. It can be seen that for non-vanishing extrinsic curvature and/or $\Lambda$ we have that Eq. (\ref{5}) is non-linear in $\psi$, making it extremely difficult to solve in general. However, it can be seen that if we take 
\begin{equation}
K^2 = 3 \Lambda \qquad {\rm and} \qquad A_{ij}=0
\end{equation}
then Eq. (\ref{6}) is satisfied identically, and Eq. (\ref{5}) becomes linear in $\psi$:
\be
\label{helmholtz}
8\tilde{D}^2 \psi = \psi \tilde{\mathcal{R}} \, .
\ee
If $\tilde{\mathcal{R}}$ is constant, then this is simply the Helmholtz equation. If $\tilde{\mathcal{R}}=0$ then it is just the Laplace equation. In either case, it has known solutions. What is more, due to the linearity of Eq. (\ref{helmholtz}), we can add any number of particular solutions to obtain new solutions. This makes it highly amenable to the study of the many-body problem in cosmology.

For the study that follows, we take the conformal metric of the initial hypersurface to be a 3-sphere of constant curvature, with line-element
\begin{equation}
\label{le}
     ds^2 = \psi^4(d\chi^2 + \sin^2{\chi}(d\theta^2 + \sin^2{\theta}d\phi^2)) \, .
\end{equation}
This means that $\tilde{R}=6$, and that solutions exist to Eq. (\ref{helmholtz}) of the form $\psi \propto 1/{\sin \left({\chi}/{2} \right)}$. Due to the linearity of Eq. (\ref{helmholtz}), this means that we can sum arbitrarily many such solutions to obtain
\begin{equation}
\label{9}
     \psi(\chi, \,\theta, \,\phi) = \sum_{i=1}^N \frac{\sqrt{\tilde{m}_i}}{2f_i(\chi,\,\theta,\,\phi)} \, .
\end{equation}
As each term in this sum corresponds to a different mass, we have a solution that corresponds to $N$ masses on a 3-sphere. The parameters $\tilde{m_i}$ are a set of constants that we refer to as the ``mass parameters'', and the $f_i$ are a set of source functions of the form $f_i = {\sin \left({\chi_i}/{2} \right)}$, where $\chi_i$ is the $\chi$ coordinate after a rotation so that the $i$th mass appears at the coordinate position $\chi=0$. This is exactly the same intrinsic geometry that one obtains for the corresponding situation on a time-symmetric hypersurface in the absence of $\Lambda$ (see \cite{tim2} for further details). 

\section{Proper Mass of Black Holes}\label{sec:proper}

It should be noted that while the $\tilde{m_i}$ in Eq. (\ref{9}) look like mass parameters, they are not the masses that one would determine from observing gravitational interactions from within the spacetime. This is due to the existence of interaction energies, which themselves gravitate and must be accounted for in order to calculate the total proper mass of each of the black holes \cite{mass}. In this section we outline the method we use to extract the proper mass of each of the black holes in our models.

The basic idea here is to look at the asymptotic form of the geometry given in Eq. (\ref{9}) in the limit $\chi_i \rightarrow 0$, where we have rotated the coordinate system so that the $i$th mass is located at $\chi_i =0$. Taking this limit corresponds to looking at the black hole from infinity, in the asymptotically flat region on the far side of the Einstein-Rosen bridge (for an explanation of this interpretation in terms of reflection operators and embedding diagrams, see \cite{reflection}). Once we know the leading-order part of the gravitational field in this limit, we can compare it to the same limit of the Schwarzschild-de Sitter (or Kottler) solution, and read off a value for the mass parameter. This process works, as we expect the geometry of space to approach that of a suitably chosen slice through the Schwarzschild-de Sitter solution as we approach $\chi_i =0$.

The Schwarzschild-de Sitter spacetime has a line-element, in standard Schwarzschild coordinates $(t, \,r,\, \theta,\, \phi)$, that can be written as
\begin{equation}
\label{SdS}
     ds^2 = -\left(1-\frac{2M}{r}-\frac{\Lambda r^2}{3}\right)dt^2 + \frac{dr^2}{\left(1-\frac{2M}{r}-\frac{\Lambda r^2}{3}\right)} + r^2(d\theta^2 + \sin^2{\theta}d\phi^2) \, ,
\end{equation}
where $M$ is the mass of the black hole. This solution has both black hole and cosmological horizons, provided that the combination of parameters $M^{\,2} \Lambda$ lies within the range $0 < M^{\,2} \Lambda < 1/9$. We now need to take a slice through this spacetime that can be compared to the geometry of our initial data, as outlined in Section \ref{sec:solution}.

The initial data constructed in Section \ref{sec:solution} has constant mean curvature (CMC), which means we need to look for a CMC foliation of the Schwarzschild-de Sitter spacetime in order to determine the mass of our black holes. In order to do this it is convenient to write the metric as \cite{estabrook} 
\begin{equation}
     ds^2 = -\left(\alpha^2 -\frac{\psi}{\beta^2} \right) d\tilde{t}^2 + 2\beta d\tilde{t}dr + \psi dr^2 + r^2(d\theta^2 + \sin^2{\theta}d\phi^2)
\end{equation}
where $\alpha$ and $\beta$ are the lapse and shift respectively, and where $\psi=\psi(r,\tilde{t})$. Insisting on a CMC foliation, with $K=$ constant, then gives \cite{cmc}
\be
\label{cmc}
\psi^{-1} = 1-\frac{2M}{r} + \left( \sqrt{\frac{2}{3}} K  + \frac{\vert A \vert}{2}    \right)\frac{\vert A \vert}{3} r^2 \, ,
\ee
where $\vert A \vert = \sqrt{A_{ij} A^{ij}}$, and where $M$ is the mass parameter from Eq. (\ref{SdS}). The value of $\vert A \vert r^3$ in this expression is constrained to be a function of $\tilde{t}$ only, and must obey the following evolution equation
\be
\frac{d (\vert A \vert r^3)}{d \tilde{t}} = \sqrt{6} \alpha M - \frac{\sqrt{6} r^2}{\psi} \frac{\partial \alpha}{\partial r} + \left( \frac{2 \vert A \vert}{\sqrt{6}} - \frac{K}{3} \right) \alpha \vert A \vert r^3 \, .
\ee
For further details about this foliation, including explicit forms for the shift and lapse functions, the reader is referred to Ref. \cite{cmc}.

The particular leaf we require is the one on which $A_{ij}=0$. From Eq. (\ref{cmc}), this gives us a hypersurface with intrinsic geometry
\be
ds^2 = \left( 1+ \frac{M}{2 \rho} \right)^4 (d\rho^2 + \rho^2 (d\theta^2 + \sin^2 \theta d \phi^2)) \, ,
\ee
where we have transformed to an isotropic radial coordinate using $r = \rho (1+ M/2 \rho)^2$. This is manifestly the same intrinsic geometry as a time-symmetric slice though the Schwarzschild geometry. Given that the intrinsic geometry in Eqs. (\ref{le}) and (\ref{9}) is also identical to that of a time-symmetric slice through the Schwarschild solution, this means that the proper mass of the $i$th black hole must be given by \cite{reflection}
\be
\label{15}
M_i = \sum_{j \neq i} \frac{\sqrt{\tilde{m}_i \tilde{m}_j}}{\sin \left( \frac{\chi_{ij}}{2} \right) },
\ee
where the indices $i$ and $j$ label each of the $N$ masses in our cosmological model, and where $\chi_{ij}$ is the coordinate distance $\chi$ between the mass $m_i$ and the mass $m_j$, after rotating so that one of these is at $\chi = 0$.

The sum in Eq. (\ref{15}) is over all other masses in the spacetime, and therefore has a total of $N-1$ terms. This means that the mass of any given black hole in this model depends on the location and mass parameter of every other black hole, through $\chi_{ij}$ and $\tilde{m}_j$, respectively. One cannot, therefore, determine the properties of any region of space without knowing the positions and magnitudes of every other mass in the universe.

\section{Back-Reaction Effect on the Deceleration Parameter}\label{sec:decel}

We are now in a position to be able to calculate some of the differences in the large-scale behaviour of a regular lattice of black holes in the presence of $\Lambda$, and a homogeneous and isotropic FLRW model containing dust and $\Lambda$. One of the most relevant and interesting quantities for a comparison of this kind is the deceleration parameter, $q$, which is a dimensionless quantity that measures the rate of change of expansion of space. For an FLRW universe, the deceleration parameter is defined as 
\begin{equation}
\label{q1}
    q \equiv - \frac{a\ddot{a}}{\dot{a}^2}
\end{equation}
where overdots denote derivatives with respect to the proper time of comoving observers, $t$. The Friedmann equations for an FLRW universe containing pressureless dust and $\Lambda$, which give us the values of $\dot{a}$ and $\ddot{a}$ in terms of the matter content, are given by 
\begin{eqnarray}
\label{17}
    \frac{\dot{a}^2}{a^2} &= \frac{8 \pi \rho}{3} -\frac{k}{a^2} + \frac{\Lambda}{3} \\\label{18}
     \frac{\ddot{a}}{a} &= -\frac{4 \pi \rho}{3} + \frac{\Lambda}{3} \, ,
\end{eqnarray}
where $k$ is the spatial curvature constant and $\rho$ is the energy density in the dust. 

The situation we considered in Section \ref{sec:solution}, where we constructed our initial data, was one in which $K^2 = 3 \Lambda$ and where $8 \tilde{D}^2 \psi = \psi \tilde{\mathcal{R}}$. The most closely analogous FLRW solutions are therefore those that obey the conditions
\begin{equation}
\label{moment}
    \frac{\dot{a}^2}{a^2} = \frac{\Lambda}{3} \qquad {\rm and} \qquad \frac{8 \pi \rho}{3} =\frac{k}{a^2} \, .
\end{equation}
This is because $K=-3 \dot{a}/{a}$ in an FLRW solution, because $\tilde{\mathcal{R}} = 6 k$, and because the electric part of the Weyl tensor (proportional to $\psi^{-1} \tilde{D}^2 \psi$) plays the role of the energy density in the effective Friedmann equations \cite{tim1}. Substituting these conditions into Eq. (\ref{q1}) then gives
\begin{equation}
\label{q2}
q = \frac{4 \pi \rho}{\Lambda} -1 \, ,
\end{equation}
which is the expression we will use for the deceleration in both our black hole lattice and our comparison FLRW space. In the former of these cases the energy density, $\rho$, will be taken to be the sum of proper masses of all black holes, and in the latter case it will simply be the energy density in dust at the moment described in Eq. (\ref{moment}). Of course, there also exist many other ways that one could construct measures of deceleration in inhomogeneous universes. The one presented in Eq. (\ref{q2}) should therefore be considered a choice, albeit one that allows for a particularly simple generalisation of the corresponding quantity in FLRW cosmology.

A quantity that can now be used as a measure of back-reaction in our black hole lattice models is the ratio of deceleration parameters between these two types of universe, which according to the discussion above is given by
\begin{equation}
\label{19}
\frac{q_L}{q_F} = \frac{4 \pi \rho_L-\Lambda}{4 \pi \rho_F-\Lambda} \, ,
\end{equation}
where subscripts $L$ and $F$ on a quantity denote that it is associated with either the lattice of black holes or the fluid of dust, respectively. One may note that for vanishing $\Lambda$ the first term on the right-hand side of Eq. (\ref{q2}) diverges, and so this quantity cannot be defined in that case. For non-zero $\Lambda$, however, the deceleration parameter is finite, and the quantity in Eq. (\ref{19}) is well defined.

In order to fully specify $\rho_F$ we now need to impose one further condition, which essentially corresponds to choosing which of the infinite family of solutions that obey the conditions Eq. (\ref{moment}) we wish to compare to our lattice universe. We choose this condition to be such that the total mass of the dust fluid is equal to the sum of the masses in the black hole lattice, as in Ref. \cite{tim2}. This uniquely specifies a single FLRW solution, and gives the energy density in the lattice and fluid models as 
\begin{equation}
\label{rhos}
\rho_L = \frac{M_T}{2 \pi^2} \frac{1}{a_L^{3}} \qquad {\rm and} \qquad    \rho_F = \frac{M_T}{2 \pi^2} \frac{1}{a_F^{3}} \, ,
\end{equation}
where $M_T =\sum_i M_i$ is the total mass in the universe, where $a_L$ and $a_F$ denote the global scale factors in the lattice and fluid models. In each case we have taken the volume of a hypersurface of constant $t$ to be given by $V=2 \pi^2 a^{3}$. It now remains to identify an appropriate scale factor for each of the two models.

In an FLRW model the choice of scale factor is unique and unambiguous. For a universe with positive spatial curvature ($k=+1$), at the moment specified by Eq. (\ref{moment}) and for the energy density specified in Eq. (\ref{rhos}), we have
\begin{equation}
\label{af}
a_F = \frac{4 M_T}{3 \pi} \, .
\end{equation}
The corresponding quantity in the lattice model is more difficult to identify, as the length of any given curve in the space depends on its location and orientation (the space is both inhomogeneous and anisotropic). We choose to take the curves that constitute the edges of the lattice cells in order to define a scale factor in this case. These curves are uniquely determined by the symmetries of the lattice, and are maximally far away from every black hole, making them as close as possible to a measure of the scale of the cosmology. The scale of these curves, as a ratio of curves that subtend the same angle at the centre of the conformal 3-sphere, is given in Table \ref{tab:table1}. These ratios are the same as in the absence of $\Lambda$ \cite{tim2}, as the intrinsic geometry in Eqs. (\ref{le}) and (\ref{9}) is unchanged from that case. Together with Eq. (\ref{af}), these ratios determine the scale factor $a_{L}$ for each of the six possible tessellations of the 3-sphere.

\begin{table}[h!]
  \centering
  \begin{tabular}{|c|c|}
    \hline
Number of masses, $N$    & Ratio of scale factors, $a_L/a_F$ \\ \hline
 5  &  $1.360$  \\ \hline
8  &  $1.248$   \\ \hline
16  &  $1.097$  \\ \hline
24  &  $1.099$\\ \hline
120  & $1.034$ \\\hline
600  & $1.002$ \\ \hline 
  \end{tabular}
  \caption{Ratios of scale factors for discrete and continuous universes for each of the six possible lattice configurations on a 3-sphere \cite{tim2}.}
   \label{tab:table1}
\end{table}
We now have all the information needed to calculate the ratio of deceleration parameters in Eq. (\ref{19}), as a function of the combination $\Lambda M_T^2$. It can be seen that in the limit $\Lambda M_T^2 \rightarrow \infty$, both $q_L$ and $q_F$ tend to $-1$. For small values of $\Lambda$ the deceleration parameters $q_L$ and $q_F$ are both large and positive (diverging in the limit $\Lambda \rightarrow 0$, as discussed above). These properties are true for each of the six tessellations of the 3-sphere, and in each case we find that $q \rightarrow -1$ rapidly as $\Lambda$ is increased. This is non-surprising, and is due to the form of the expression in Eq. (\ref{19}). 

\begin{figure}[t]
    \centering
        \includegraphics[width=1
        \textwidth]{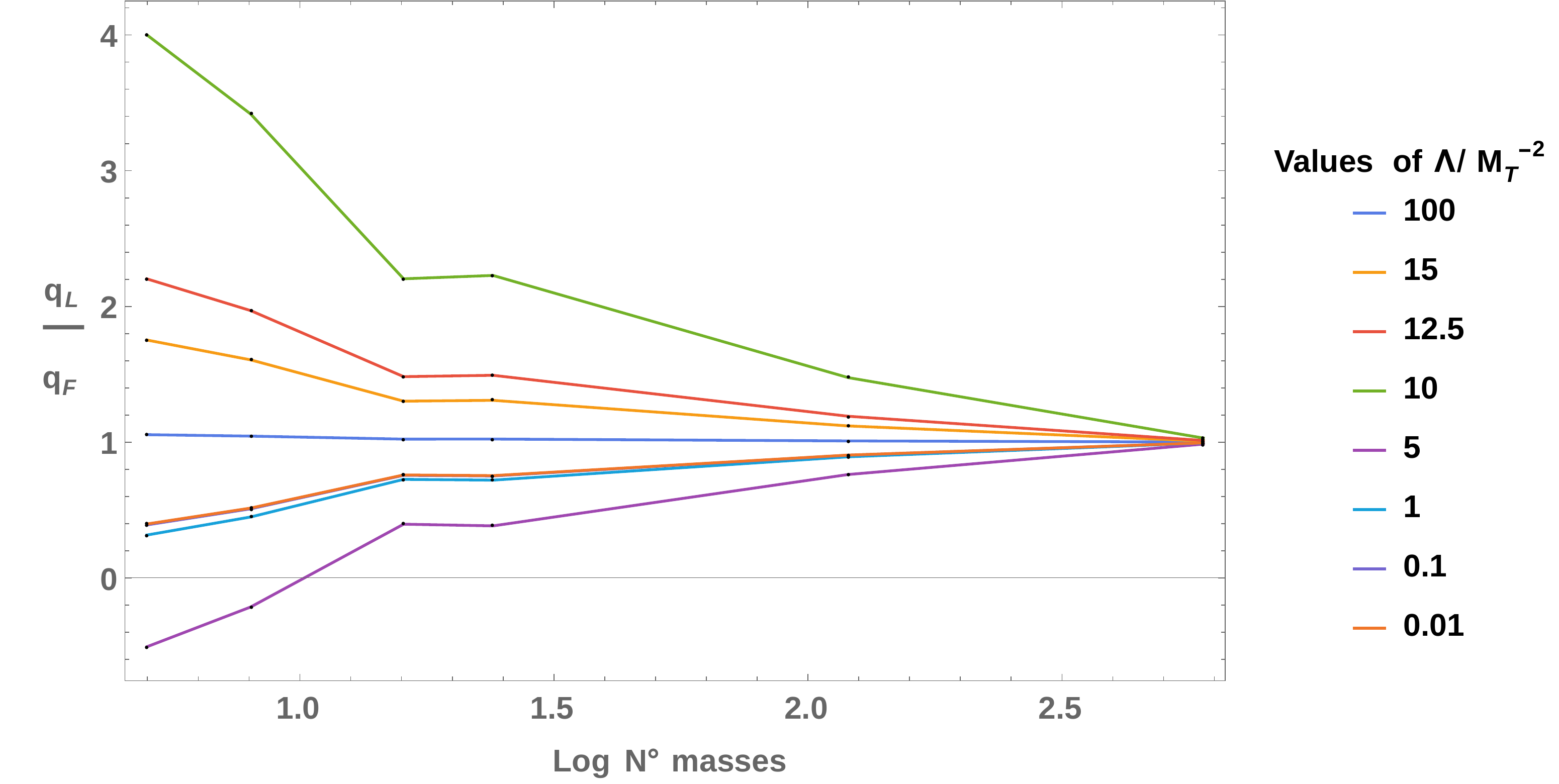}
        \caption{The ratio of the deceleration parameter in the lattice universe, compared to the corresponding FLRW universe, in each of the six possible tessellations of the 3-sphere. Each curve corresponds to a different value of $\Lambda$, in units of $M_T^{-2}$.}
        \label{fig:fig1}
\end{figure}

%For large values of $N$ the scale factors increase, whilst $\rho \propto a^{-3}$ results in $\rho$ decreasing. Using $q = 4 \pi \rho \Lambda^{-1} - 1$, for large values of $N$ and hence small values of $\rho$, the second term dominates and we have $q \approx -1$. 
Fig. \ref{fig:fig1} shows the ratio of our two deceleration parameters, evaluated at different values of $\Lambda$. As the number of masses increases the ratios tend to unity for all values of $\Lambda$. This is expected as for large values of $N$ we have $a_L \approx a_F$, as can be seen from Table \ref{tab:table1} as well as from studies that consider very large numbers of randomly located masses \cite{newref}. This equality of scale factors implies $\rho_L \approx \rho_F$, and therefore $q_L/q_F \approx 1$. When $\Lambda$ is large, as in the $100 M_T^{-2}$ case, the ratio of deceleration parameters is approximately unity regardless of $N$. Again, this is to be expected from the form of the expression in Eq. (\ref{19}). When $\Lambda$ is very small, the ratio of deceleration parameters reduces to $\rho_L / \rho_F = (a_F/a_L)^3$. This is shown by the curves corresponding to $0.1 M_T^{-2}$ and $0.01 M_T^{-2}$, which are indistinguishable from each other. On the other hand, for intermediate values of $\Lambda$, when $4 \pi \rho_F = \Lambda$, the same ratio diverges. Using Eqs. (\ref{rhos}) and (\ref{af}) this occurs at $\Lambda \approx 8.33 M_T^{-2}$, for which the green curve is a good indicator of this divergence in ratio. All values of $\Lambda$ below this critical value give $q_L/q_F < 1$, while all values above it give $q_L/q_F >1$.

\section{Locating the Apparent Horizons}\label{sec:locations}

We now wish to determine the effect a non-zero cosmological constant has on the locations of horizons in each of our lattice models. The apparent horizons in this case are defined as closed 2-surfaces that have outgoing future-directed normal null geodesics with vanishing expansion. Such surfaces are marginally outer trapped, meaning that light rays are neither able to escape to infinity nor fall towards the singularity \cite{book}. To find the positions of these surfaces we use the orthonormal frame approach \cite{orth}, with non-zero $\Lambda$  but in the absence of matter, and follow the approach used in Ref. \cite{tim1}.

We first consider a unit time-like vector $u_{\mu}$ and its covariant derivative, decomposed as 
\begin{equation}
    \nabla_{\mu}u_{\nu} = -u_{\mu}\dot{u_{\nu}} + \sigma_{\mu\nu} + \frac{1}{3}\Theta h_{\mu\nu} - \omega_{\mu\nu},
\end{equation}
where $\Theta$ is the expansion scalar, $\sigma_{\mu\nu}$ is the shear tensor, $\omega_{\mu\nu}$ is the vorticity tensor and $h_{\mu\nu}=g_{\mu \nu}+ u_{\mu} u_{\nu}$ is the projection tensor. If $u^{\mu}$ is orthogonal to our initial hypersurface then we immediately have
\be
\label{25}
K=- \Theta, \qquad A_{ij} = \frac{dx^{\mu}}{d\xi^i}\frac{dx^{\nu}}{d\xi^j} \sigma_{\mu \nu}, \qquad {\rm and} \qquad \omega_{\mu \nu}=0 \, ,
\ee
where $\xi^i$ are coordinates in this space. 
%A generalised Friedmann equation from \cite{orth} is 
%\be
%0 = -\frac{1}{3}\Theta^2 - \frac{1}{2}{}^\star R + \Lambda, 
%\ee
%where we immediately see that $\star R$, the Ricci scalar of the hypersurface, vanishes using Eq. (\ref{25}). 
We now need to consider the electric part of the Weyl tensor. This quantity is obtained by taking the trace free part of the Riemann tensor, $R_{\mu\nu}\,\,^{\rho\sigma}$, and decomposing it into its electric and magnetic parts \cite{orth}. We find that for the solution outlined in Section \ref{sec:solution}, and using the results in Eq. (\ref{25}),  the electric part of the Weyl tensor can be written as
\begin{equation}
\label{er}
     E_{\mu\nu} = \mathcal{R}_{\mu\nu},
\end{equation}
where $\mathcal{R}_{\mu\nu}$ is the Ricci tensor of the initial hypersurface. This result is obtained using the general expressions for the Gauss embedding equation in orthonormal frame variables \cite{orth}, and is identical to the result that one obtains in the absence of $\Lambda$ \cite{tim1}. We will now use it to locate the position of the apparent horizons.

Consider three orthogonal space-like unit vectors, $\{ e_1^{\,\mu}, \,e_2^{\,\mu}, \,e_3^{\,\mu} \}$ arranged so that $e_1^{\,\mu}$ points outwards from the black hole. These three vectors, along with $u^\mu = e_0^{\,\mu}$, complete our orthonormal frame. We can now write the condition that outgoing null geodesics $k^{\mu}$ should have zero expansion as \cite{book}
\begin{equation}
\label{27}
\hat{\Theta} \equiv  k_{\mu ; \nu} m^{\mu \nu} = \frac{1}{\sqrt{2}} \left( e_{0\mu;\nu} m^{\mu\nu} + e_{1\mu;\nu} m^{\mu\nu} \right) = 0,
\end{equation}
where $m^{\mu\nu}$ is the screen space projector, which is an induced two-dimensional metric that can be written as $m^{\mu\nu} = e_2^{\,\mu} e_2^{\,\nu} + e_3^{\,\mu} e_3^{\,\nu}$. The first term in Eq. (\ref{27}) is directly related to the extrinsic curvature of the initial hypersurface, and does not vanish if $\Lambda$ is non-zero. The second term in Eq. (\ref{27}) is the expansion of $e_1^{\,\mu}$ in the screen space. 

We treat the black holes horizons as being non-expanding in the initial data, so that $\hat{\Theta} = \mathcal{L}_k \hat{\Theta} =0$. The null Raychaudhuri equation then tells us that the shear of the integral curves of $k^{\mu}$ must also vanish, which means that the positions of the apparent horizons in the initial data must have constant mean curvature and hence must be totally geodesic with indeterminate lines of curvature \cite{book2}.  These properties mean that the space-like normal to the apparent horizons must be a principal direction of the Ricci tensor of the 3-space \cite{book3}. From Eq. (\ref{er}) this means that we have $E_{12}=E_{13}=0$ at all points on the apparent horizon, and hence that \cite{orth}
\begin{equation}
\label{eE}
     {\bf e}_1(E^{11}) = 3 a_1 E^{11} + n_{23}(E^3_{\; 3} - E^2_{\; 2}) \, ,
\end{equation}
where ${\bf e}_1$ is a frame derivative, $2 a_1 =  -e_{1\mu;\nu} m^{\mu\nu}$ is the expansion of $e_1^{\,\mu}$, $n_{23}$ is a symmetric 2-index object and $E^{11}, E^3_{\; 3}$ and $E^2_{\; 2}$ are components of the electric part of the Weyl tensor. Eq. (\ref{eE}) can be used to find the positions of the horizons in our initial data.

Let us now rotate our lattice of black holes so that one of the masses appears at $\chi=0$, and consider the points at which the horizon of this black hole intersects curves that exhibit local rotational symmetry (such as the edges of cells, or the curves that connect it to neighbouring black holes). At these points we have $E^3_{\; 3} = E^2_{\; 2}$ \cite{tim1}, which means that the second term on the right-hand side of Eq. (\ref{eE}) vanishes. For the value of $a_1$ we find 
\begin{eqnarray}
 a_1^{\rm \,outer} = - \frac{1}{2}\,e_{1\mu;\nu} m^{\mu\nu} = \frac{1}{2}\,e_{0\mu;\nu} m^{\mu\nu} = \frac{1}{3}\Theta = \pm \,\sqrt{\frac{\Lambda}{3}} \, ,
\end{eqnarray}
where we have used Eq. (\ref{27}) in the second equality, and where in the third equality we have used the fact that $m^{\mu\nu}$ contains only two of the three orthonormal basis vectors. The $\pm$ after the final equality indicates the fact that our initial data can describe either an expanding ($+$) or a collapsing ($-$) space.

We can repeat the analysis above for marginally inner trapped surfaces, which will also be of interest in the models we are constructing. In this case Eq. (\ref{27}) should be modified so that it instead gives $e_{0\mu;\nu} m^{\mu\nu} - e_{1\mu;\nu} m^{\mu\nu}= 0$. By exactly the same logic, this yields
\begin{eqnarray}
     a_1^{\rm \,inner} =- \frac{1}{2}\,e_{1\mu;\nu} m^{\mu\nu}= -\frac{1}{2}\,e_{0\mu;\nu} m^{\mu\nu}= -\frac{1}{3}\Theta = \mp \,\sqrt{\frac{\Lambda}{3}}\, ,
\end{eqnarray}
where the $\mp$ sign here corresponds to expanding ($-$) or collapsing ($+$) space. We therefore have both inner and outer trapped surfaces for each of the two possible signs of $\Theta$. Those with $\Theta < 0$ (or $K>0$) correspond to contracting universes whilst those with $\Theta > 0$ (or $K<0$) correspond to expanding universes. 

\begin{figure}[t]
\label{penrose}
\scalebox{0.5}{
\begin{tikzpicture}
\node (I)    at ( 4,0)   {};
\node (1)    at ( 7.8,1.2)   {};
\node (2)    at ( 7.8,-1.2)   {};
\node (II)   at (-4,0)   {};
\node (III)  at (0, 2.5) {};
\node (IV)   at (0,-2.5) {};
\node (V)   at (-12,0) {};
\node (VI)   at (12,0) {};
\node (VII)   at (-8,-2.5) {};
\node (3)   at (-9.1,3){$\Theta > 0$};
\node (4)   at (-9.1,-3) {$\Theta < 0$};
\node (VIII)   at (-8,2.5) {};
\node (IX)   at (8,-2.5) {};
\node (X)   at (8,2.5) {};

\path  % Four corners of left diamond
  (II) +(90:4)  coordinate  (IItop)
       +(-90:4) coordinate (IIbot)
       +(0:4)   coordinate                  (IIright)
       +(180:4) coordinate (IIleft)
       ;
\draw (IIleft) -- 
         node[midway, above left]    {}
         node[midway, above, sloped] {OC}
     (IItop) --
         node[midway, above, sloped] {IBH}
      (IIright) -- 
          node[midway, above, sloped] {OBH}
     (IIbot) --
          node[midway, above, sloped] {IC}
          node[midway, below left]    {}    
      (IIleft) -- cycle;

\path % Four corners of the right diamond (no labels this time)
   (I) +(90:4)  coordinate (Itop)
       +(-90:4) coordinate (Ibot)
       +(180:4) coordinate (Ileft)
       +(0:4)   coordinate (Iright)
        ;

    \draw (Ibot) -- (Ileft)--(Itop); 
     \draw (Itop) -- 
     node[midway, above, sloped] {IC}
     ($(Itop)!.93!(Iright)$); 
         \draw (Ibot) --      
         node[midway, above, sloped] {OC}
         ($(Ibot)!.93!(Iright)$);
             \draw [gray]($(Itop)!.93!(Iright)$) -- (Iright); 
     \draw [gray]($(Ibot)!.93!(Iright)$) -- (Iright);

\draw  (Ileft) --
  node[midway, above, sloped] {OBH}
       (Itop);
       
     \draw  (Ileft) --
  node[midway, above, sloped] {IBH}
       (Ibot);
\path % Four corners of the right diamond (no labels this time)
   (V) +(90:4)  coordinate (Vtop)
       +(-90:4) coordinate (Vbot)
       +(180:4) coordinate (Vleft)
       +(0:4)   coordinate (Vright)
        ;
\draw (Vtop) -- (Vright) -- (Vbot) ;
% No text this time in the next diagram
\path % Four corners of the right diamond (no labels this time)
   (VI) +(90:4)  coordinate (VItop)
       +(-90:4) coordinate (VIbot)
       +(180:4) coordinate (VIleft)
       +(0:4)   coordinate (VIright)
        ;
\draw  [color=gray!100] (VItop) -- (VIleft) -- (VIbot) ;
% Squiggly lines
\draw[decorate,decoration=zigzag](IItop) -- (Itop)
      node[midway, above, inner sep=2mm] {};

\draw[decorate,decoration=zigzag](IIbot) -- (Ibot)
      node[midway, below, inner sep=2mm] {};
      
\draw(Vtop) -- (IItop)
      node[midway, above, inner sep=2mm] {};

\draw(Vbot) -- (IIbot)
      node[midway, below, inner sep=2mm] {};
      
\draw(Itop)--($(VItop)!.65!(Itop)$)
      node[midway, above, inner sep=2mm] {};

\draw(Ibot) --($(VIbot)!.65!(Ibot)$)
      node[midway, below, inner sep=2mm] {};
      
\draw[thick, dashed] 
%($(Vbot)!.5!(IIbot)$)
    (Vbot) to[out=35, in=145, looseness=1.3]
    (2);
    \draw[thick, dashed] 
%($(Vbot)!.5!(IIbot)$)
    (Vtop) to[out=-35, in=-145, looseness=1.3]
    (1);

\draw[very thick]($(VItop)!.65!(Itop)$) to[out=-60, in=60, looseness=0.8]($(VIbot)!.65!(Ibot)$) ;
\draw  [color=gray!100] (VItop) -- ($(VItop)!.65!(Itop)$) ;
\draw  [color=gray!100] (VIbot) -- ($(VIbot)!.65!(Ibot)$)  ;
     
\end{tikzpicture}}
\caption{Penrose-Carter diagram for the region of spacetime around one of the black holes, in a lattice with non-zero $\Lambda$. The hypersurfaces that constitute our initial data are shown as dashed lines. The expanding universe with $\Theta > 0$ passes through an outer trapped cosmological horizon (OC), an outer trapped black hole horizon (OBH), an inner trapped black hole horizon (IBH) and an inner trapped cosmological horizon (IC) before emerging into the cosmological region on the right. Similarly, a contracting universe with $\Theta < 0$ passes through an inner trapped cosmological horizon (IC), an inner trapped black hole horizon (IBH), an outer trapped black hole horizon (OBH) and an outer trapped cosmological horizon (OC) before emerging into the cosmology. The solid curved line on the right-hand side represents a cut-off, beyond which the causal structure of the cosmological region should be expected to be too complicated to represent in a 2D figure.}
\label{penrose}
\end{figure}
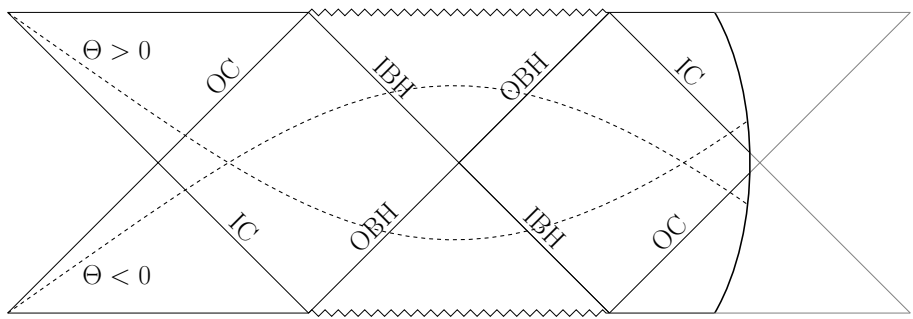

Substituting these results back into Eq. (\ref{eE}) gives us a general expression for the positions of all of the horizons in a black hole lattice universe with a cosmological constant. They are given by solutions to the following expression:
\begin{equation}
\label{31}
     {\bf e}_1(E^{11}) =  \alpha_1\alpha_2\,\sqrt{3\Lambda}\,E^{11} \, ,
\end{equation}
where we have introduced two new quantities, $\alpha_1$ and $\alpha_2$, which both take values of either $+1$ or $-1$. The first parameter, $\alpha_1$, describes whether the space is expanding or contracting, and we choose $\alpha_1 = + 1$ to correspond to expansion and $\alpha_1 = - 1$ to correspond to contraction. The second parameter, $\alpha_2$, then describes whether the horizon in question is an inner or outer trapped surface. Given our previous choice, we have that $\alpha_2 = + 1 $ corresponds to outer trapped surfaces, whilst $\alpha_2 = - 1 $ refers to inner trapped surfaces. The reader may note that this implies there is a symmetry between outer trapped surfaces in the expanding case and inner trapped surfaces in the contracting case. 

Fig. \ref{penrose} shows the hypersurfaces we have been considering, for some arbitrary value of $\Lambda$, in a Penrose-Carter diagram. We have indicated how both expanding and contracting hypersurfaces fit into this diagram. Each surface passes through exactly two cosmological horizons and two black hole horizons, and the sign of $\Theta$ determines whether these are inner or outer trapped surfaces. On the left of the diagram the spacetime approaches perfect Schwarzschild-de Sitter, while on the right it approaches the complicated cosmological region (separated off by the curve). At the mid-point, within the black hole region, each of the two spaces contains a throat that has a finite, non-zero minimal radius. In the limit $\Lambda \rightarrow 0$ we can verify that Eq. (\ref{31}) reduces to the corresponding equation in Ref. \cite{tim1}, and in that case the apparent horizon becomes degenerate with the minimal sphere that can fit within the throat at the centre of the black hole region. Finally, we note that the procedure outlined in this section correctly identifies the known locations of the horizons in the exact Schwarzschild-de Sitter geometry.

\section{Distance Between Black Holes in Regular Lattices}\label{sec:distances}

The lattice spacetime with $\Lambda$ has two types of horizon -- cosmological and black hole -- and these can be inner or outer trapped surfaces. In addition, these can occur in either contracting or expanding configurations. We may therefore naively have  thought there are eight different horizons to be found. However, noticing that an outer/inner trapped surface in the expanding case is the same as an inner/outer trapped surface in the contracting case reduces the number of horizons to be found to four. We now seek to determine the location of these horizons in each of the six possible lattices. To start with, we use the 5-mass model as an example, and rotate coordinates so that one of the masses appears at the position $\chi=0$. We then calculate the electric part of the Weyl tensor along a curve that connects this mass with one of its neighbours in an adjacent cell, as a function of radial coordinate $\chi$. This information can then be used in Eq. (\ref{31}), to obtain the positions of the various horizons for different values of $\Lambda$ in units of $M_0^{\,-2}$ (where $M_0$ is the proper mass of one of the black holes, as calculated using Eq. (\ref{15})) The results of all this are displayed in Table \ref{tab:tab2}, where we have chosen to consider an expanding universe, with $\Theta > 0$ and therefore $\alpha_1 = +1$. From this point onwards we will refer to horizons in the expanding case only, unless specified otherwise. The results and analyses for contracting solutions are exactly the same, albeit under an interchange of the words ``inner" and ``outer" when referring to horizons. 
%We restrict ourselves to looking for horizons that occur up to (and including) the midpoint between the two masses, as we wish to identify those that can be said to belong to the mass at $\chi=0$. 

For $\Lambda = 0$, we find that the locations of the horizons at $\chi=\chi_2$ and $\chi = \chi_3$ are degenerate, as previously identified in Ref. \cite{bent}. The positions of the other two horizons when $\Lambda=0$ are found to be at $\chi_1=0$ (the origin) and $\chi_4=0.912$ (the midpoint between masses). These are mathematical solutions to Eq. (\ref{31}), but for vanishing $\Lambda$ had not previously been considered as physically interesting. Indeed, although the latter can rightly be identified as an extremal surface in the geometry (it is part of the face of one of the cells in the lattice), it is not part of a closed extremal surface, and so is not technically a horizon. Similarly, it is stretching the definition somewhat to call the sphere at $\chi=0$ a horizon, as this corresponds to a sphere at infinity on the far side of the Einstein-Rosen bridge. Nevertheless, it is useful to identify these points as ``cosmological horizons'', as they become more interesting when $\Lambda \neq 0$.

\begin{table}[t]
\centering
\scalebox{0.9}{\begin{tabular}{|c|c|c|c|c|}
\hline
\multirow{2}{*}{$\Lambda  /M_0^{-2}$} & \multicolumn{2}{c|}{$\alpha_2 = +1$} & \multicolumn{2}{c|}{$\alpha_2 = -1$}\\ \cline{2-5} 
 & $\chi_1$   & $\chi_2$  & $\chi_3$  &$\chi_4$          \\ \hline
 0  &  0  &  0.413& 0.413  &   0.912\\ \hline
 $0.002$  & 0.00539   &  0.367&   0.474  &   0.824\\ \hline
$0.004$  &  0.00781   &  0.349&  0.507  &   0.781 \\ \hline
$0.006$  & 0.00975 & 0.337&   0.541   &   0.740\\ \hline
$0.008$  &  0.0115  &  0.326& 0.586   &   0.690 \\ \hline
$0.009$  & 0.0121   &  0.323&   0.638   &   0.638\\ \hline
$0.020$  &   0.0196   &  0.285   &   -   &  - \\ \hline
$0.040$  &  0.0311  &  0.241 &  - & -\\ \hline
$0.060$  & 0.0426  &  0.207  &  - & - \\ \hline
$0.080$  &   0.0559   & 0.177   & - & - \\ \hline
$0.100$  &   0.0749  &  0.144   &  - & -\\ \hline
$0.111$  &   0.104   &  0.104  &  - & -\\ \hline
$0.120$  &   -   &  -  &  - & -\\ \hline
\end{tabular}}
\vspace{10pt}
\caption{Positions of the horizons for the expanding 5-mass model, as a function of $\Lambda$ measured in units of $M_0^{-2}$. The horizons at $\chi=\chi_1$ and $\chi_2$ are both outer trapped, while those at $\chi=\chi_3$ and $\chi_4$ are both inner trapped. Dashes indicate that no horizons exist for the given value of $\Lambda$. The position of the midpoint between masses is at $\chi=0.912$, in this configuration.}
\label{tab:tab2}
\end{table}

Turning on $\Lambda$ reveals that the horizons change positions, as indicated in Table \ref{tab:tab2}. For the inner trapped surfaces with $\alpha_2 = -1$, the black hole horizon at $\chi = \chi_3$ moves outwards with increasing $\Lambda$, while the cosmological horizon at $\chi=\chi_4$ moves inwards as $\Lambda$ increases. This means that as $\Lambda$ increases, $\chi_3$ and $\chi_4$ converge towards each other, and in fact become degenerate at $\Lambda \simeq 0.009 M_0^{-2}$. For values of $\Lambda$ above this critical value, there are no solutions to Eq. (\ref{31}). Similarly, for outer trapped surfaces with $\alpha_2 = +1$, the black hole horizon at $\chi=\chi_2$ moves to lower $\chi$ as $\Lambda$ increases, whilst the cosmological horizon at $\chi=\chi_1$ moves to higher values of $\chi$. This corresponds to the black hole horizon moving outwards on the far side of the Einstein-Rosen bridge, while the cosmological horizon moves inwards from infinity. Again, these two horizons become degenerate at a critical value of $\Lambda\simeq 0.111 M_0^{-2}$, and there exist no solutions to Eq. (\ref{31}) for higher values of $\Lambda$.

\begin{figure}
    \centering
        \includegraphics[width=1\textwidth]{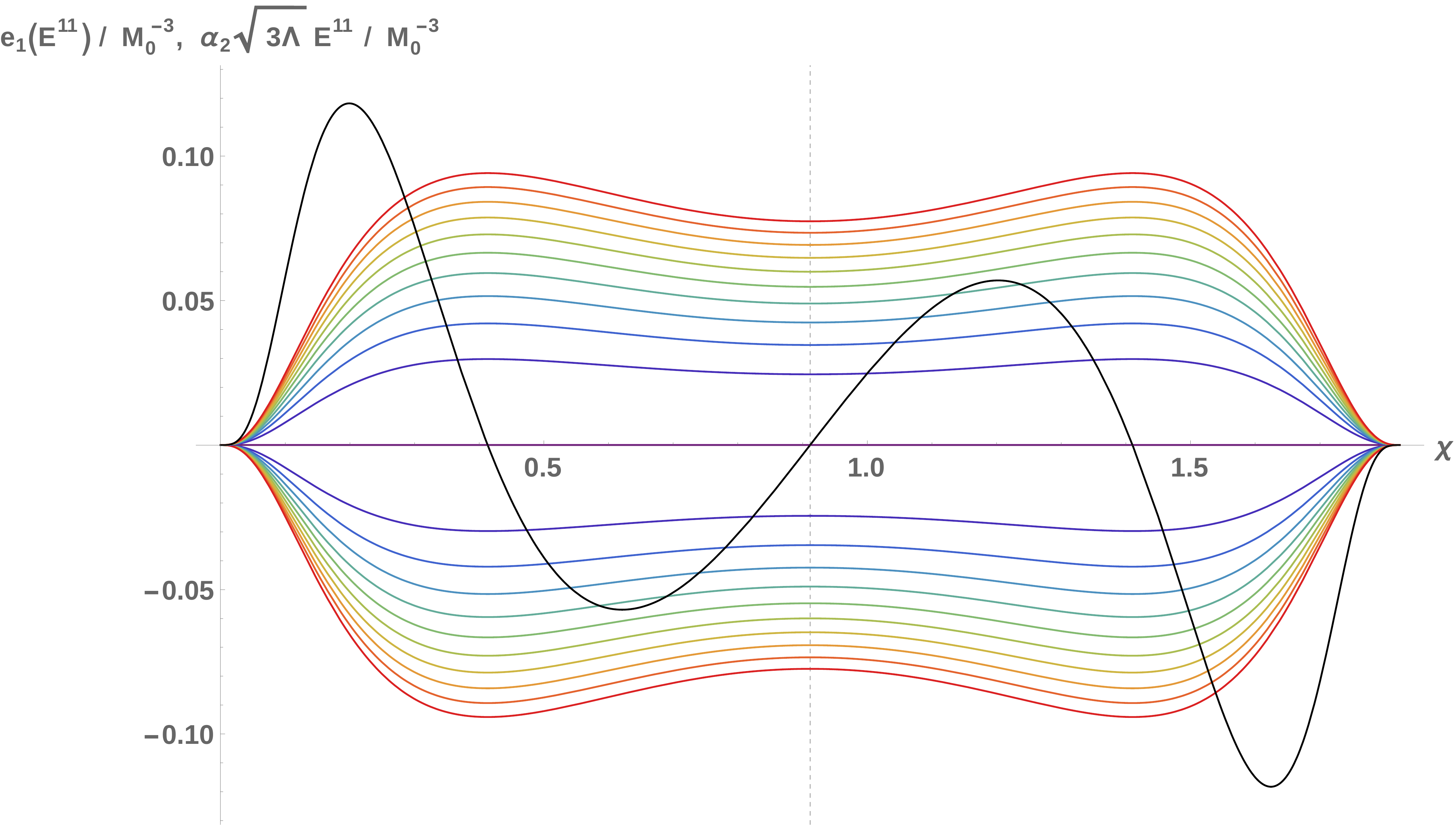}
        \caption{A graphical representation of the left and right-hand sides of Eq. (\ref{31}) for the expanding 5-mass model. The black curve corresponds to the left-hand side, while the multi-coloured curves correspond to the right-hand side. The values of $\Lambda$ range from 0 (purple) to $0.02 \,M_0^{-2}$ (red) in increments of $0.002\, M_0^{-2}$. For $\alpha_2 = + 1$ the multi-coloured lines are above the horizontal axis, and $\alpha_2 = - 1$ for those below. The vertical dashed line shows the midpoint between the two masses.}
        \label{fig:fig3}
\end{figure}

To illustrate this behaviour graphically one can consider the functional form of the left and right-hand sides of Eq. (\ref{31}). These quantities are shown in Fig. \ref{fig:fig3} as a function of the coordinate $\chi$, for the 5-mass expanding model, and with $\alpha_2 = \pm 1$. The coloured lines represent the right-hand side of Eq. (\ref{31}), and hence are a function of $\Lambda$. When the black line and the coloured lines cross, Eq. (\ref{31}) is satisfied, and a value for the position of a horizon, $\chi_h$, can be read off. For small enough values of $\Lambda$ there are four occasions where this happens. However, if the value of $\Lambda$ is increased sufficiently then these lines do not cross at all, and the horizons cease to exist.

\begin{figure}
\vspace{30pt}
    \centering
    \begin{subfigure}[b]{0.49\textwidth}
        \includegraphics[width=\textwidth]{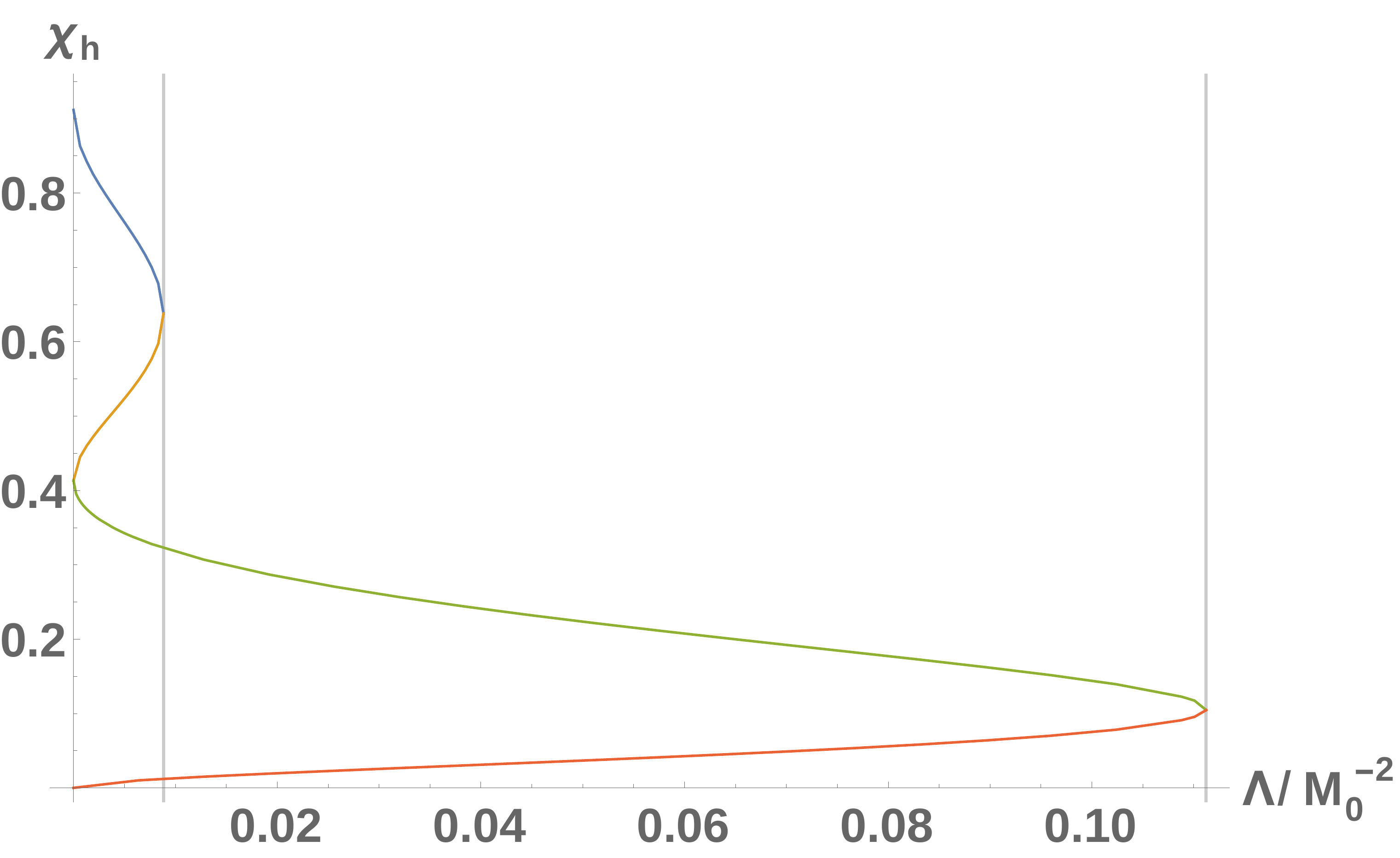}
        \caption{\centering{5 masses.}}
    \end{subfigure}
    \vspace{5mm}
    \begin{subfigure}[b]{0.49\textwidth}
        \includegraphics[width=\textwidth]{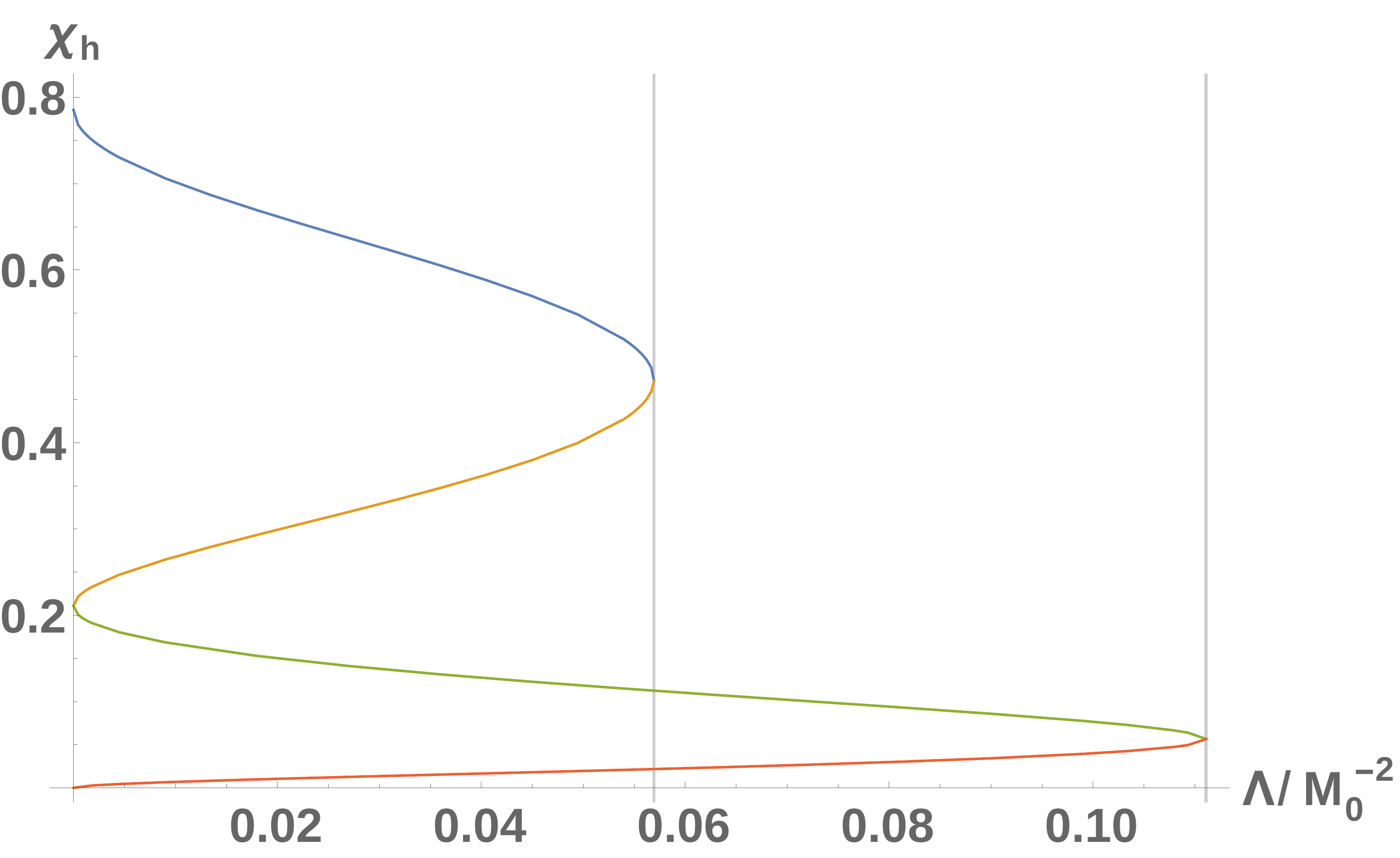}
           \caption{\centering{8 masses.}}
    \end{subfigure}
    \begin{subfigure}[b]{0.49\textwidth}
        \includegraphics[width=\textwidth]{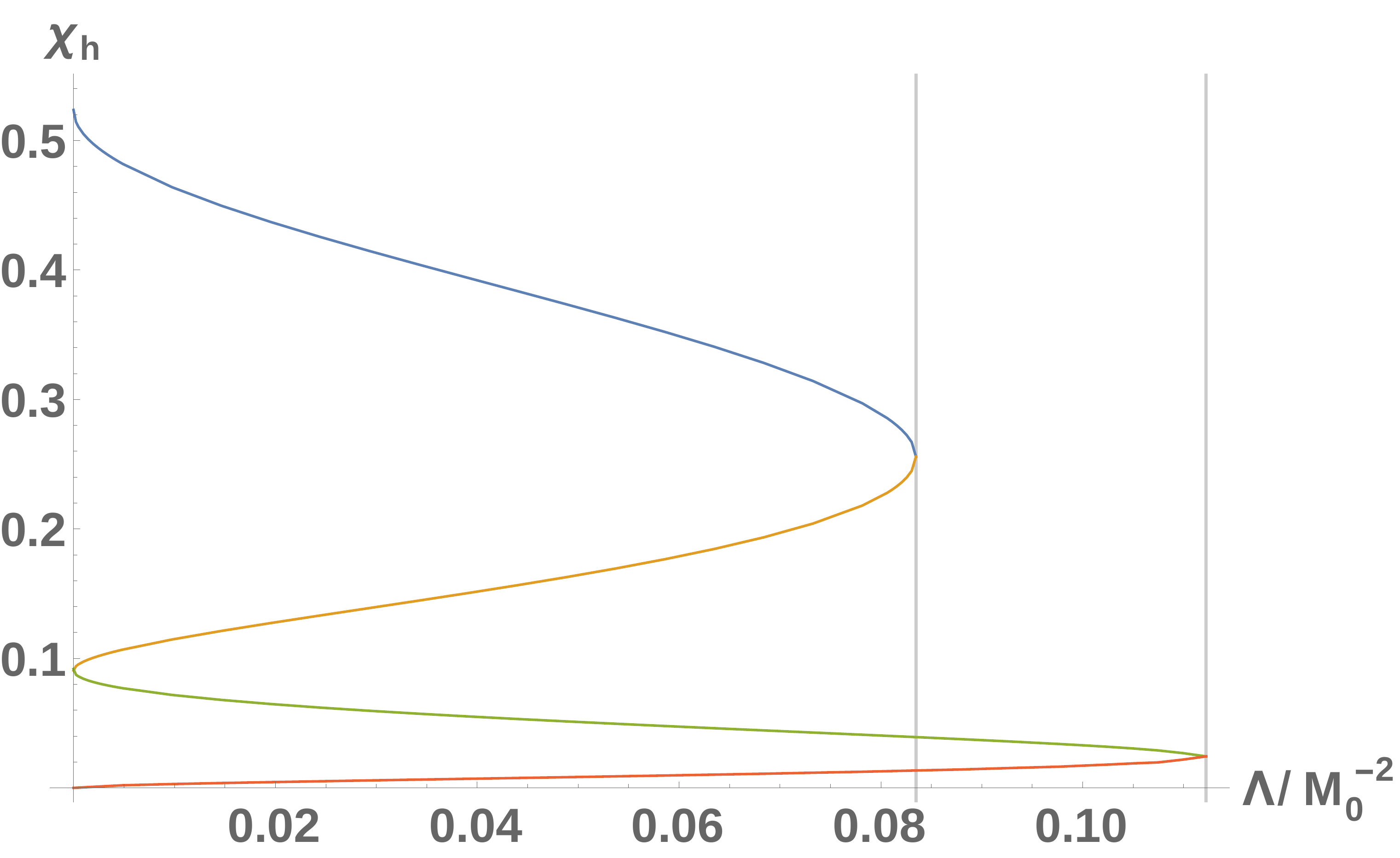}
           \caption{\centering{16 masses.}}
    \end{subfigure}
    \vspace{5mm}
    \begin{subfigure}[b]{0.49\textwidth}
        \includegraphics[width=\textwidth]{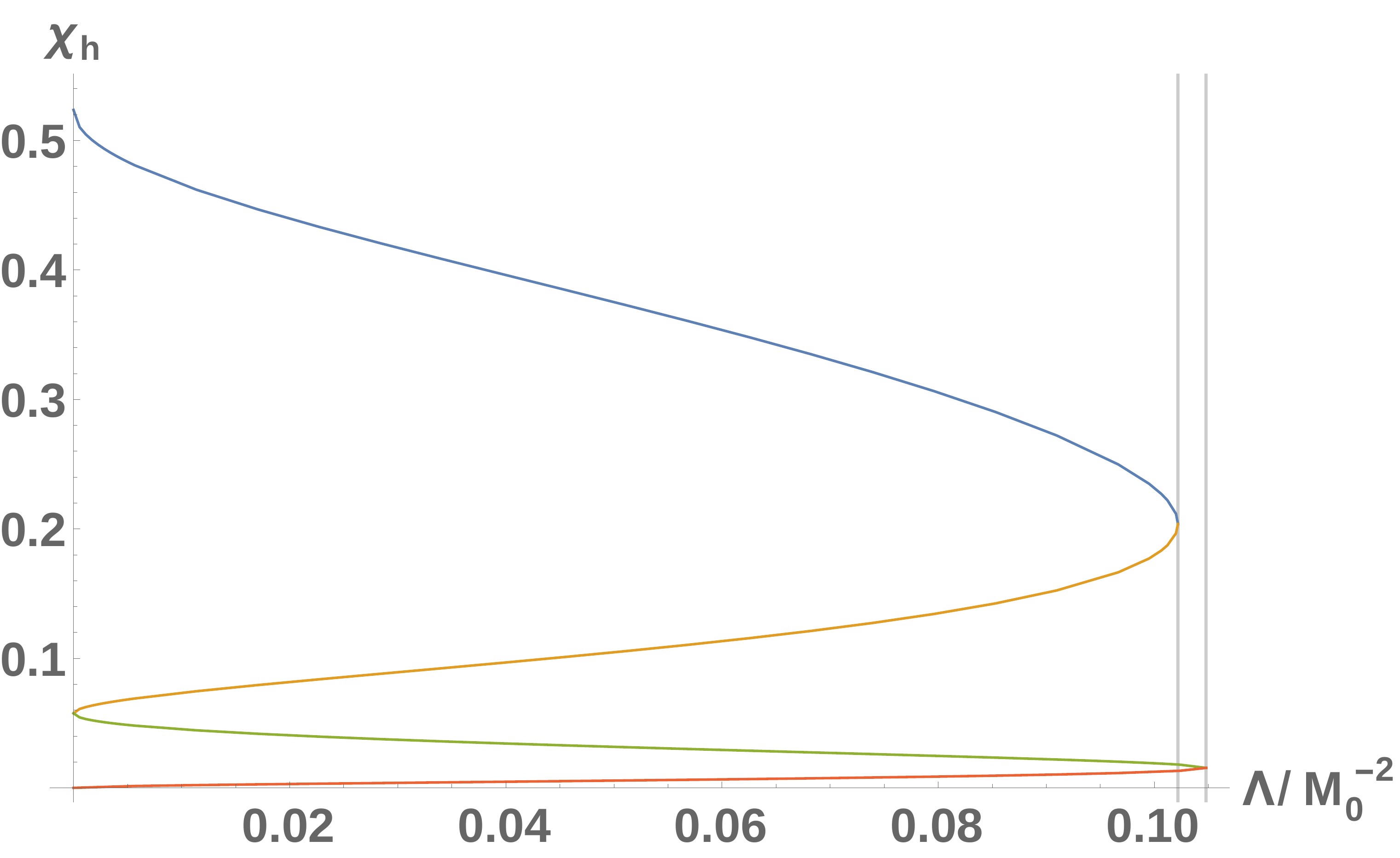}
           \caption{\centering{24 masses.}}
    \end{subfigure}
    \begin{subfigure}[b]{0.49\textwidth}
        \includegraphics[width=\textwidth]{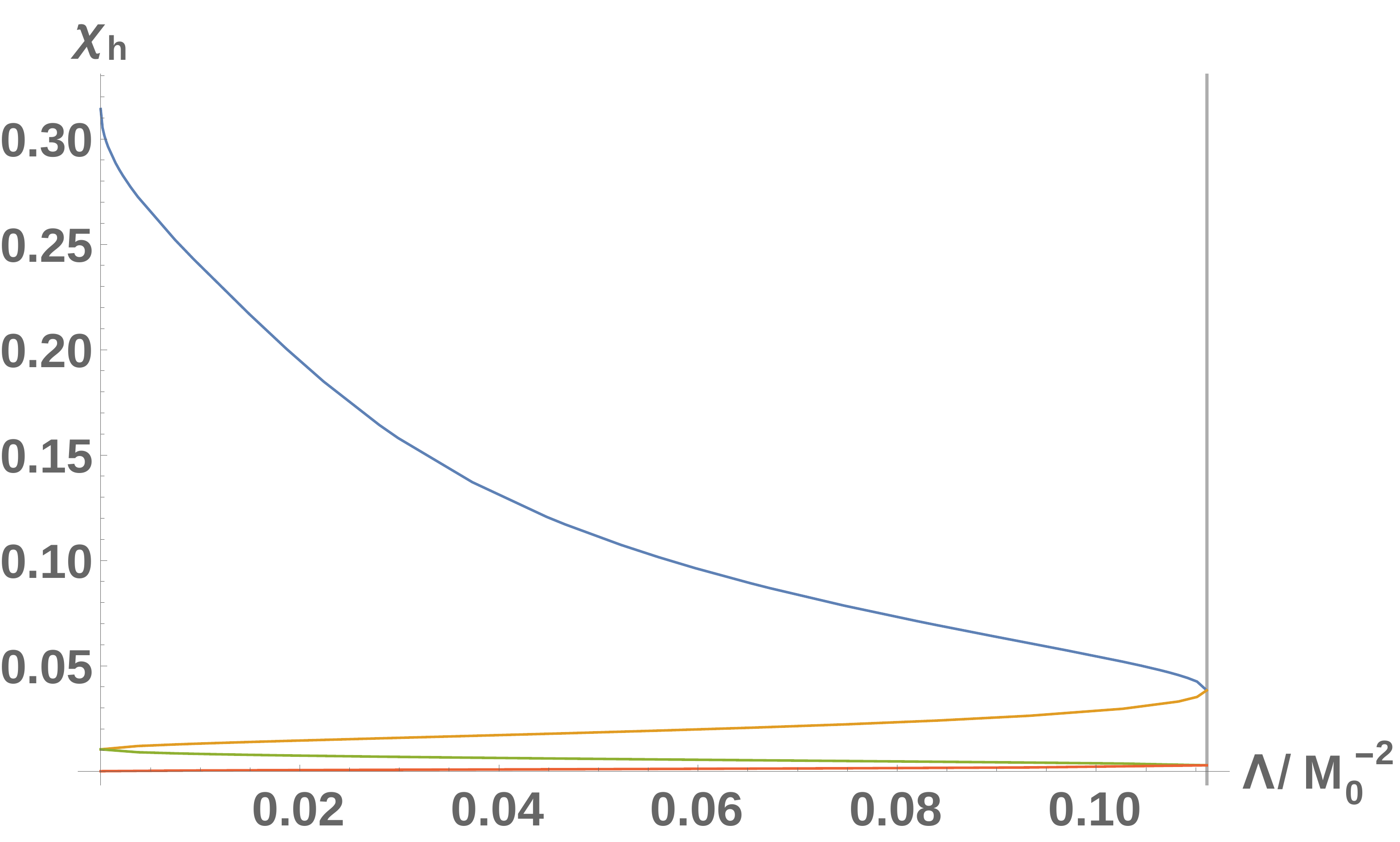}
           \caption{\centering{120 masses.}}
    \end{subfigure}
      \begin{subfigure}[b]{0.49\textwidth}
        \includegraphics[width=\textwidth]{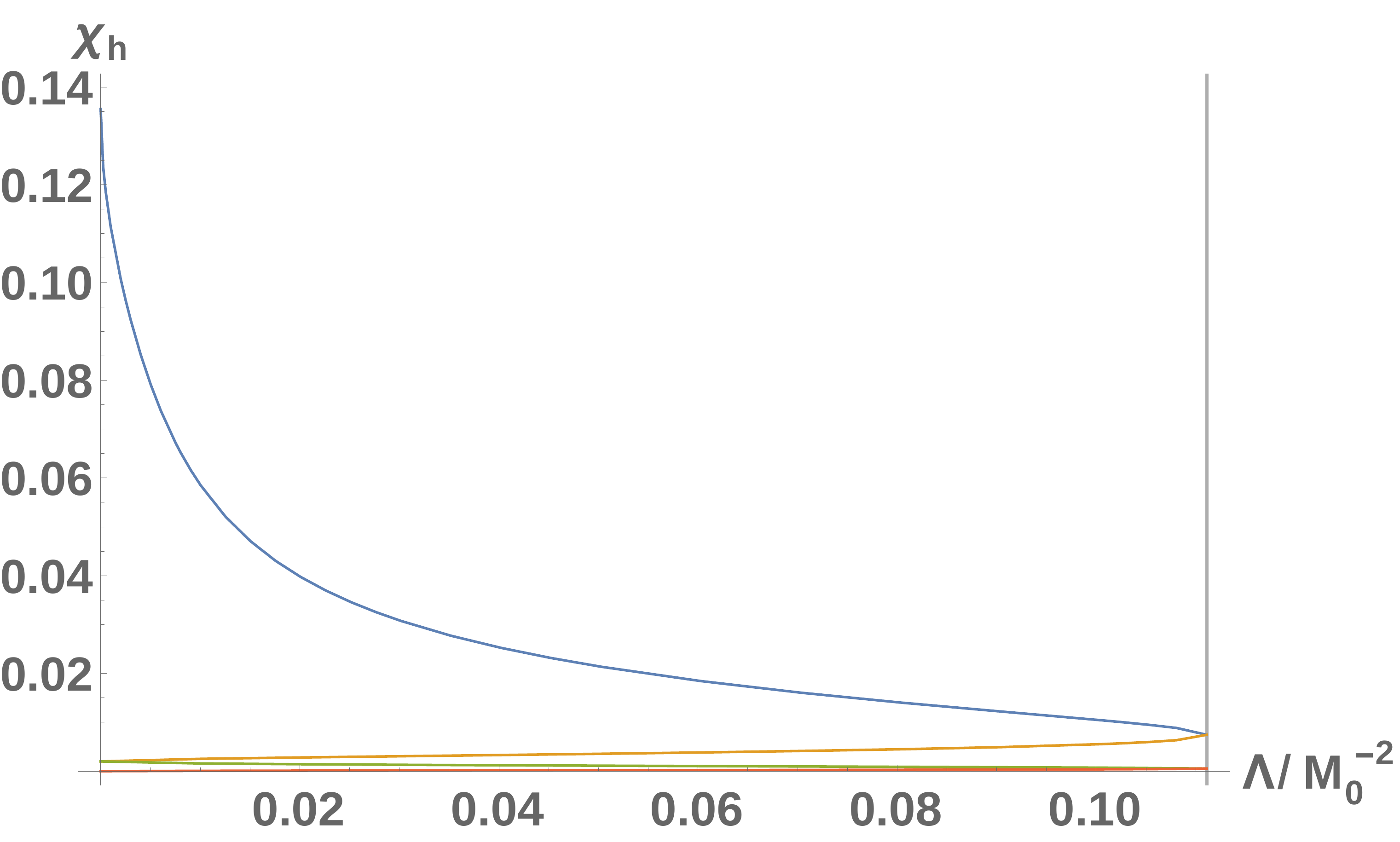}
           \caption{\centering{600 masses.}}
    \end{subfigure}
    \vspace{10pt}
    \caption{The positions of the four possible horizons in each of the six lattice universes, as a function of $\Lambda$. The blue (orange) line corresponds to a marginally inner trapped cosmological (black hole) horizon, whilst the green (red) line corresponds to a marginally outer trapped black hole (cosmological) horizon. The vertical lines represent the value of $\Lambda$ after which no solutions to Eq. (\ref{31}) exist.}
    \label{fig:fig4}
\end{figure}

The $\chi$ values from the positions of the horizons can be plotted separately, and in the first panel of Fig. \ref{fig:fig4} these values are shown for the 5-mass model. The subsequent panels in Fig. \ref{fig:fig4} show the positions of the horizons for each of the five other lattice universes that we are considering, for different values of $\Lambda$. These diagrams show all four possible horizons, both inner and outer black hole horizons as well as inner and outer cosmological horizons. The extremal values of $\Lambda$, in both the inner and outer regions, are displayed in the diagrams as vertical lines. As the number of masses is increased, these two extremal values for $\Lambda$ converge, until they become indistinguishable by eye in the case of the 120-mass and 600-mass configurations. The black hole horizons meet at $\Lambda=0$, as expected \cite{bent}, and in every case the largest $\chi_4$ value is simply the midway point between the two masses. The information presented in Fig. \ref{fig:fig4} can be used to determine the distance between neighbouring black holes for every value of $\Lambda$, in each of the six possible configurations.

 \section{Extremal Values for $\Lambda$}\label{sec:extremal}

\begin{figure}
    \centering
        \includegraphics[width=0.9\textwidth]{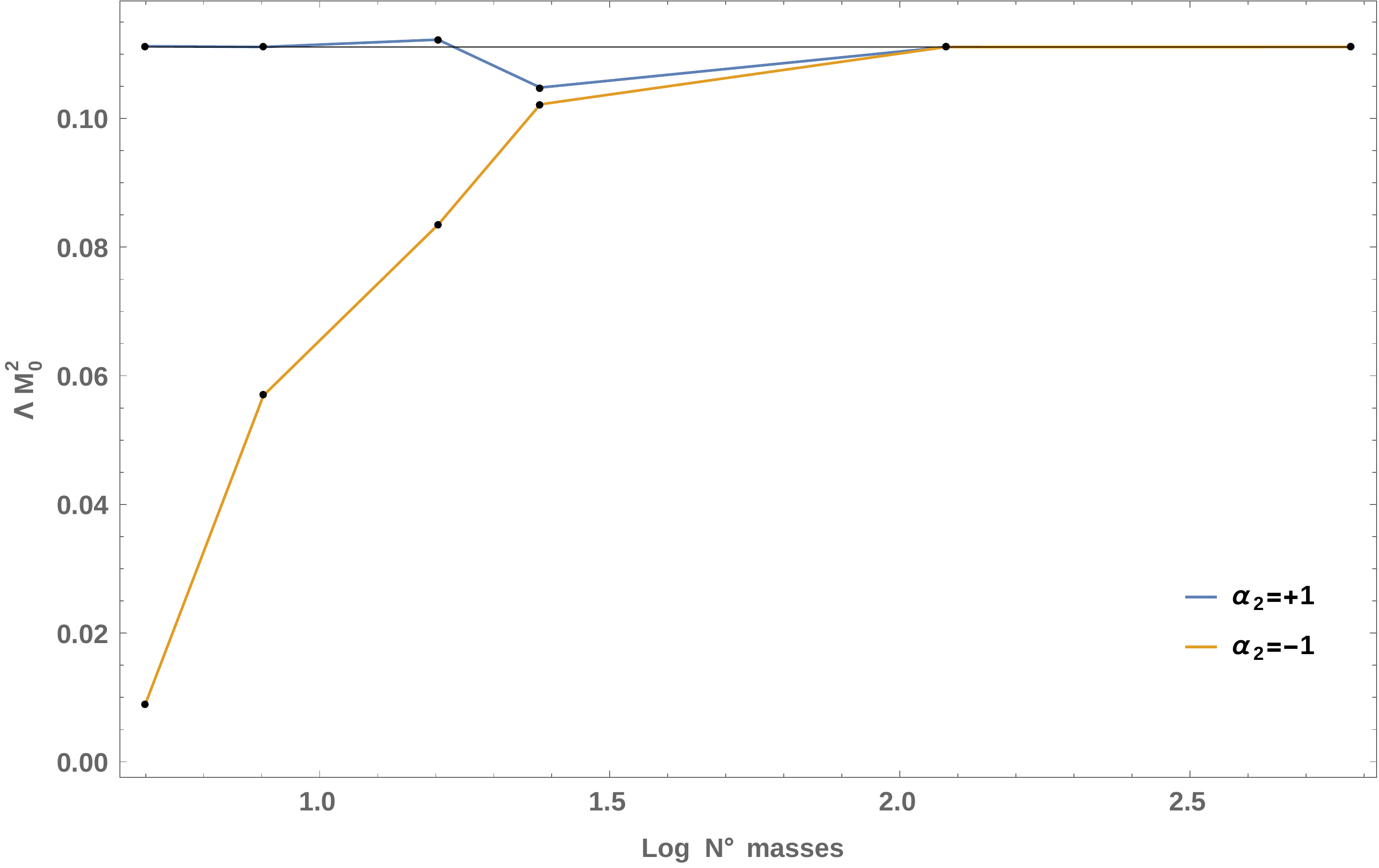}
        \caption{The upper bound on $\Lambda M_0^{\,2}$ as a function of the number of masses, for the outer horizons (blue) and the inner horizons (yellow). The solid line corresponds to the Schwarzschild-de Sitter value of ${1}/{9}$.}
        \label{fig:fig5}
\end{figure}

Cosmological and black hole horizons exist in the Schwarschild-de Sitter spacetime provided $0 < M_0^{\,2} \Lambda < 1/9$. In our lattice models there also exist bounds on the combination $M_0^{\,2} \Lambda$, but the value of the upper limit is not always equal to $1/9$. This is an interesting result, as it appears that the fact the black holes exist within a cosmology changes the value of $M_0$ for which the black holes can be said to be extremal (for a given value of $\Lambda$). The vertical lines in Fig. \ref{fig:fig4} have already been used to denote the location of the point where the black hole and cosmological horizons become degenerate. Here we will collect these results, for each of the six lattice models, to consider the behaviour of the upper bound on $M_0^{\,2} \Lambda$ as a function of the number of masses in the universe. We present this information numerically in Table \ref{tab:tab3}, and illustrate it graphically in Fig. \ref{fig:fig5}. It is manifest that two different phenomena arise as the number of masses increases. The first is that the critical values for $\Lambda$, after which there are no horizons, converge to the same value for both $\alpha_2 = +1$ and $\alpha_2=-1$. The second is that the value to which they converge is the same value as the upper bound in the Schwarschild-de Sitter solution,  $M_0^{\,2} \Lambda = 1/9$. This is not unexpected, as for large values of $N$ the distance between neighbouring masses increases, and the overall contribution from any individual black hole to the spacetime of any other diminishes. In other words, increasing $N$ isolates each black hole to the point that it can be very well approximated by the Schwarzschild-de Sitter solution.

It is also interesting to see if a varying lower limit for $\Lambda M_0^2$ exists, or whether it is always bound from below by zero. One way of determining this is to look at the gradients of both sides of Eq. (\ref{31}) and see if it always true that $\partial_\chi(\sqrt{3\Lambda}E^{11}) > \partial_\chi(\psi^{-2}\partial_\chi(E^{11}))$, for arbitrarily small values of $\Lambda$. If it is, then the inner black hole and cosmological horizons should be expected to exist, and to remain separate. We calculated these two functions numerically, and found the condition to be valid down to values of $\Lambda \sim \mathcal{O}(10^{-14})$ and $\chi \sim \mathcal{O}(10^{-8})$ for all six lattice models. This indicates the lower bound on $\Lambda M_0^2$ is always zero, at least as far as can be determined with machine precision calculations. This result agrees with the results of Ref. \cite{yoo}, who found that an inner cosmological horizon always exists in the case of an infinite flat lattice.

\begin{table}
\centering
\scalebox{0.9}{\begin{tabular}{|c|c|c|}
\hline
\multirow{2}{*}{Number of masses, $N$} & \multicolumn{2}{c|}{$\Lambda / M_0^{-2}$} \\
\cline{2-3} 
 & $\alpha_2 = -1$   & $\alpha_2 = +1$ \\ \hline
 5  &  $0.00884$  &  $0.111$ \\ \hline
8  &  $0.0569$  &  $0.111$  \\ \hline
16  &  $0.0835$  &  $0.112$ \\ \hline
24  &  $0.102$  &  $0.105$ \\ \hline
120  & $0.111$  &  $0.111$ \\ \hline
600  & $0.111$  &  $0.111$ \\ \hline
\end{tabular}}
\caption{Extremal values for $\Lambda$, for each of the six expanding lattice models. The column headed $\alpha_2=-1$ gives the upper bound on $\Lambda M_0^2$ for the inner horizons, while the corresponding quantity for the outer horizons is given in the column headed $\alpha_2=+1$.}
\label{tab:tab3}
\end{table}

\section{Discussion}\label{sec:discuss}

Inhomogeneous cosmological models are important for understanding the role of small-scale structures on the large-scale dynamics of the universe, and in turn the consequent effects that this has on observable quantities. This paper contributes to the literature on this subject by providing and studying exact initial data for a universe that contains regularly arranged black holes in a hyperspherical universe and in the presence of a cosmological constant. In Section \ref{sec:solution} we presented the constraint equations for such a universe, and solved them in closed analytic form. In Section \ref{sec:proper} we then determined the proper mass of the black holes by considering the constant mean curvature foliation of the Schwarzschild-de Sitter spacetime. 

We then studied the change in deceleration parameter due to inhomogeneity in spaces of this type, by comparing to positively curved FLRW models that contain the same cosmological constant and dust with the same total proper mass. The results of this were presented in Section \ref{sec:decel}. We found that the back-reaction effect from inhomogeneity decreases as the number of masses is increased, and as the value of the cosmological constant becomes large. We then found expressions for the locations of the black hole and cosmological horizons in Section \ref{sec:locations}, and determined numerical values for the positions of these horizons in each of the six possible lattice universes in Section \ref{sec:distances}. We found that there are upper bounds on the value of $\Lambda M_0^2$, where $M_0$ is the proper mass of the black holes in this spacetime, and that this upper bound converges to the Schwarschild-de Sitter value as the number of masses in the universe is increased. 

These results give qualitative agreement with the numerical work found performed in Ref. \cite{yoo} when considering the inner black hole and cosmological horizons, but appear to disagree when considering the outer horizons. We expect that the source of this disagreement is due to the freedom that exists to change the scale of the size of the cells in the flat lattice that the authors of this study had been considering. No such freedom exists in the hyperspherical lattices that we have studied, which makes them much more rigid, and thus changes the relationship between the positions of black hole horizons and cosmological horizons.

The initial data we have constructed and studied can be evolved numerically, as was done for the 8-mass case with vanishing $\Lambda$ in Ref. \cite{bent}. In fact, it may be that the present set of initial data has certain benefits over the corresponding set without $\Lambda$. This is because the initial hypersurface when $\Lambda=0$ is time-symmetric, meaning that evolving the initial data in either direction in time means evolving it towards a cosmological singularity, where numerical errors are likely to increase. Initial data for eternally expanding lattice universes has so far only been achieved numerically \cite{yoo,yoo0,bent2}, which inevitably leads to numerical errors that can grow during the evolution. In our case, we strongly suspect that, for the initial data we presented in Section \ref{sec:solution}, there will be values of the parameter combination $\Lambda M_0^2$ that lead to eternally expanding universes. The fact that the initial data is exact in this case may then mean that errors in the future evolution are easier to control. Using the Friedmann equations, and the values given in Eq. (\ref{moment}), we estimate the required conditions for eternal expansion will be approximately when $\Omega_{\Lambda} \gtrsim \frac{4}{27} \Omega_m$, where we have defined $\Omega_\Lambda \equiv \Lambda/3H_0^{\,2}$ and $\Omega_m \equiv 8\pi \rho_m/3H_0^{\,2}$.

\section*{Acknowledgements}

We would like to thank Reza Tavakol, Kjell
Rosquist and Eloisa Bentivegna for helpful discussions. JD and TC both acknowledge support from the STFC.


\section*{References}
\begin{thebibliography}{10}

\bibitem{wiltshire}D.~Wiltshire, What is dust? Physical foundations of the averaging problem in cosmology, \textit{Class. Quant. Grav.} \textbf{28}, 164006 (2011).

\bibitem{linder}E.~V.~Linder, Resource letter: Dark energy and the accelerating universe, \textit{Am. J. Phys.} \textbf{76}, 197 (2008).

\bibitem{buchert}T.~Buchert, On average properties of inhomogeneous fluids in general relativity. 1. Dust cosmologies, \textit{Gen. Rel. Grav.} \textbf{32}, 105 (2000).

\bibitem{chris}C.~Clarkson, G.~Ellis, J.~Larena and O.~Umeh, Does the growth of structure affect our dynamical models of the universe? The averaging, backreaction and fitting problems in cosmology, \textit{Rept. Prog. Phys.} \textbf{74} 112901 (2011).

\bibitem{kras}K.~Bolejko, M.-N.~Celerier and A.~Krasinski, Inhomogeneous cosmological models: Exact solutions and their applications, \textit{Class. Quant. Grav.} \textbf{28}, 164002 (2011).

\bibitem{lind}R.~W.~Lindquist and J.~A.~Wheeler, Dynamics of a lattice universe by the Schwarzschild-cell method, \textit{Rev. Mod. Phys.} \textbf{29}, 432 (1957).

\bibitem{mis}C.~W.~Misner, The method of images in geometrostatics, \textit{Ann. Phys.} \textbf{24}, 102 (1963).

\bibitem{tim2}T.~Clifton, K.~Rosquist and R.~Tavakol, An exact quantification of backreaction in relativistic cosmology, \textit{Phys. Rev.} \textbf{D86}, 043506 (2012).

\bibitem{bent}E.~Bentivegna and M.~Korzy\'{n}ski, Evolution of a periodic eight-black-hole lattice in numerical relativity, \textit{Class. Quant. Grav.} \textbf{29}, 165007 (2012).

\bibitem{ev1}C.~M.~Yoo, H.~Okawa and K.-i.~Nakao, Black hole universe: Time evolution, 
{\it Phys. Rev. Lett.} {\bf 111}, 161102 (2013).

\bibitem{ev2}E.~Bentivegna and M.~Korzy\'{n}ski, Evolution of a family of expanding cubic black-hole lattices in numerical relativity, {\it Class. Quant. Grav.}, {\bf 30},  235008 (2013).

\bibitem{tim1}T.~Clifton, D.~Gregoris, K.~Rosquist and R.~Tavakol, Exact evolution of discrete relativistic cosmological models, \textit{JCAP} \textbf{1311}, 010 (2013).

\bibitem{yoo}C.-M.~Yoo and H.~Okawa, Black hole universe with a cosmological constant, \textit{Phys. Rev.} \textbf{D89}, 123502 (2014).

\bibitem{ev4}M.~Korzy\'{n}ski, I.~Hinder and E.~Bentivegna, On the vacuum Einstein equations along curves with a discrete local rotational and reflection symmetry, {\it JCAP} {\bf 08}, 025 (2015).

\bibitem{ev5}T.~Clifton, D.~Gregoris and K.~Rosquist, The magnetic part of the Weyl tensor, and the expansion of discrete universes, arxiv:160700775 (2016).

\bibitem{brune}J.-P.~Bruneton and J.~Larena, Dynamics of a lattice Universe: The dust approximation in cosmology, \textit{Class. Quant. Grav.} \textbf{29}, 155001 (2012).

\bibitem{av1}T.~Clifton, Cosmology without averaging, {\it Class. Quant. Grav.} {\bf 28}, 164011 (2011).

\bibitem{av2}V.~A.~A.~Sanghai and T.~Clifton, Post-Newtonian cosmological modelling, {\it Phys. Rev. D} {\bf 91}, 103532 (2015); erratum {\it Phys. Rev. D} {\bf 93}, 089903 (2016).

\bibitem{vij}V.~A.~A.~Sanghai and T.~Clifton, Cosmological backreaction in the presence of radiation and a cosmological constant, \textit{Phys.Rev.} \textbf{D94}, 023505 (2016).

\bibitem{mass}T.~Clifton, What's the matter in cosmology?, arXiv:1509.06682 (2015).

\bibitem{reflection}T.~Clifton, The method of images in cosmology, \textit{Class. Quant. Grav.} \textbf{31}, 175010 (2014).

\bibitem{estabrook}F.~Estabrook, H.~Wahlquist, S.~Christensen, B.~DeWitt, L.~Smarr and E.~Tsiang,  Maximally slicing a black hole, \textit{Phys. Rev.} \textbf{D7},  2814 (1973).

\bibitem{cmc}K.-I.~Nakao, K.-I.~Maeda, T.~Nakamura and K.-I.~Oohara, The
constant-mean-curvature slicing of the Schwarzschild-de Sitter space-time,  \textit{Phys. Rev.} \textbf{D44}, 1326 (1991).

\bibitem{newref}M.~Korzy\'{n}ski, Backreaction and continuum limit in a closed universe filled with black holes,  \textit{Class. Quant. Grav.} \textbf{31}, 085002 (2014).

\bibitem{book}T.~W.~Baumgarte and S.~L.~Shapiro, Numerical Relativity, Cambridge University Press (2010).

\bibitem{orth}H.~v. Elst and C.~Uggla, General relativistic $1+3$ orthonormal frame approach revisited, \textit{Class. Quant. Grav.} \textbf{14}, 2673 (1997).

\bibitem{book2}L.~P.~Eisenhart, Riemannian Geometry, Princeton University Press (1925).

\bibitem{book3}D.~J.~Struik, Grundz$\ddot{{\rm u}}$ge der Mehrdimensionalen Differentialgeometrie iu Direkter Darstellung, Springer Berlin (1922).

\bibitem{yoo0}C.-M.~Yoo, H.~Abe, Y.~Takamori and K.-i.~Nakao, Black Hole Universe: Construction and Analysis of Initial Data, {\it Phys. Rev.} {\bf D86}, 044027 (2012).

\bibitem{bent2}E.~Bentivegna, Solving the Einstein constraints in periodic spaces with a multigrid approach, {\it Class. Quant. Grav.} {\bf 31}, 035004 (2014).

\end{thebibliography}
\end{document}